\documentclass[aps,prl,twocolumn,superscriptaddress]{revtex4-2}
\usepackage{graphicx}
\usepackage{color}

\newcommand{\km}{$k$-means}

\usepackage{mhchem}

\begin{document}
\title{Electron dynamics in a three-dimensional Brillouin zone analysed by machine learning}
\author{Paulina Majchrzak}
\affiliation{Department of Physics and Astronomy, Aarhus University, 8000 Aarhus C, Denmark}
\author{Charlotte Sanders}
\affiliation{Central Laser Facility, STFC Rutherford Appleton Laboratory, OX11 0QX, Harwell, UK}
\author{Yu Zhang}
\affiliation{Central Laser Facility, STFC Rutherford Appleton Laboratory, OX11 0QX, Harwell, UK}
\author{Andrii Kuibarov}
\affiliation{Leibniz IFW Dresden, 01069, Dresden, Germany}
\author{Oleksandr Suvorov}
\affiliation{Leibniz IFW Dresden, 01069, Dresden, Germany}
\author{Emma Springate}
\affiliation{Central Laser Facility, STFC Rutherford Appleton Laboratory, OX11 0QX, Harwell, UK}
\author{Iryna Kovalchuk}
\affiliation{Leibniz IFW Dresden, 01069, Dresden, Germany}
\affiliation{Kyiv Academic University, 03142 Kyiv, Ukraine}
\author{Saicharan Aswartham}
\affiliation{Leibniz IFW Dresden, 01069, Dresden, Germany}
\author{Grigory Shipunov}
\affiliation{Leibniz IFW Dresden, 01069, Dresden, Germany}
\author{Bernd B\"uchner}
\affiliation{Leibniz IFW Dresden, 01069, Dresden, Germany}
\author{Alexander N. Yaresko}
\affiliation{Max-Planck-Institute for Solid State Research, 70569, Stuttgart, Germany}
\author{Sergey Borisenko}
\affiliation{Leibniz IFW Dresden, 01069, Dresden, Germany}
\author{Philip Hofmann}
\email{philip@phys.au.dk}
\affiliation{Department of Physics and Astronomy, Aarhus University, 8000 Aarhus C, Denmark}

\begin{abstract}
The electron dynamics in the unoccupied states of the Weyl semimetal PtBi$_2$ is studied by time- and angle-resolved photoemission spectroscopy (TR-ARPES). The measurement's result is the photoemission intensity $I$ as a function of at least four parameters: the emission angle and kinetic energy of the photoelectrons, the time delay between pump and probe laser pulses, and the probe laser photon energy that needs to be varied to access the full three-dimensional Brillouin zone of the material. The TR-ARPES results are reported in an accompanying paper \cite{short_paper}. Here we focus on the technique of using \km, an unsupervised machine learning technique, in order to discover trends in the four-dimensional data sets. We study how to  compare the electron dynamics across the entire data set and how to reveal subtle variations between different data sets collected in the vicinity of the bulk Weyl points.
\end{abstract}
\maketitle

\section{Introduction}

Experimental physics generates ever larger data sets, creating challenges for data analysis and storage. Traditionally, the problem is most pronounced in particle physics and astronomy but rapid technical progress has lead to high data volumes and data generation rates in other fields, such as condensed matter physics. Even for relatively small data sets, an interpretation can be challenging when the data is multi-dimensional. As an example of this, we discuss the ultrafast electron dynamics in the Weyl semimetal (WSM) PtBi$_2$ in the accompanying paper \cite{short_paper}. Despite of a data set size on the order of only hundreds of megabytes, the measured quantity depends on four variables, making it difficult  to establish qualitative trends.

A versatile tool for discovering such trends are clustering approaches, in particular \km, an unsupervised machine learning technique \cite{MacQueen:1967aa,Ball:1967aa}. The \km~ algorithm can be used to sort data into a pre-defined number of $k$ clusters based on similarity, giving easy access to trends in the data. \km~ can potentially reveal hidden patterns and the technique has a very wide range of applications, from image compression to classification of large data sets in astronomy or particle physics \cite{Bock:2007wz}.

The particular setting here is a measurement of the electron dynamics in PtBi$_2$ by time- and angle-resolved photoemission spectroscopy (TR-ARPES); for a review on the technique see \cite{Boschini:2024aa}. The quantity of interest is the photoemission intensity, $I$, as a function of electron energy, $E$, measured with respect to the Fermi energy $E_F$, one emission angle (or the wave vector parallel to the surface in one direction $k$), delay time  between the excitation by an ultrashort pump laser pulse and the measurement by a second extreme UV laser pulse $\Delta t$, and photon energy of that laser pulse, $h\nu$. The dependence of $I$ on four parameters $(E, k, \Delta t, h\nu)$ introduces several difficulties that are not encountered in most conventional ARPES experiments: (1) It is hard to discover systematic trends in the multi-dimensional parameter space. (2) Data reduction is challenging. In conventional ARPES, one can often fit the data by simple models. For instance, energy distribution curves and especially momentum distribution curves can be often approximated by simple functions, even in the presence of many-body effects \cite{Valla:1999ab,Hofmann:2009ab}. In time-resolved ARPES, this does not necessarily work as we shall see. In particular the dependence of $I$ on $\Delta t$ cannot usually be described by a simple line shape model throughout the data set. 

The main objective of the analysis discussed in this paper is to unravel the electron dynamics in different parts of the three-dimensional (3D) bulk Brillouin zone (BZ). The location of an ARPES measurement within the BZ is mainly given by $k$ and $h\nu$ and we thus focus on the time dependence $I(\Delta t)$ for different values of  $(k, h\nu)$ and energy $E$. In particular, we introduce the photoemission time distribution curve (TDC), $I(\Delta t)$, for fixed values of $(E, k, h\nu)$. Examples of TDCs are seen in Fig. \ref{lfig:1}(b),(d) and (f). TDCs typically show a fast excitation of electrons by the pump laser at around $\Delta t = 0$, followed by a decay encoding the different channels available for the excited electrons to loose energy. It is not always possible to come up with a simple line shape model -- for example a  single exponential  -- for the decay part of the TDC throughout the data set, challenging our ability to establish a unified understanding of the electron dynamics in the material.

In order to address this situation, we explore different approaches to cluster TDCs by \km. Using this tool, we can identify regions of the BZ with faster or slower decay times, regions with similar decay line shapes across different values of $(E, k, h\nu)$, as well as subtle differences in the decay for the same $(E, k)$ when $h\nu$ is changed. The strengths of \km~in the present context are: (1) We do not have to compare parameters of a fit to the TDCs, such as a decay time, but we can compare the TDCs directly without having to make assumptions about their specific line shapes. (2) A product of the clustering are the cluster centroids, the averaged TDC line shape over the entire cluster. These centroids have a much higher signal to noise ratio ratio ($S/N$) than the individual TDCs. (3) We can apply \km~either to the entire data set or to combinations of TDCs from different photon energies, revealing overall trends or specific differences between photon energies. 

The paper is structured as follows: We first explore how to best apply \km~ to the type of data we are handling here, using data from a single photon energy $h\nu$. This underlines the benefits and drawbacks of the \km-based analysis. We then apply \km~to the entire data set composed of spectra collected at three different values of $h\nu$. We conclude the paper by summarising the main results, focusing on the use of \km~to reveal subtle trends in a multi-dimensional data set.

\section{Data from a single photon energy}

Our eventual goal is to use \km~ in order to understand trends in the data measured at different photon energies, throughout the 3D BZ. However, it is instructive to first explore the strengths and limitations of clustering by using a data set from a single photon energy. To this end, we illustrate different implementations of \km~ clustering on the data set obtained at 27.4~eV. This corresponds to a $k$-space cut close to the Weyl points (WPs) in the 3D BZ (see accompanying paper Ref. \cite{short_paper}). 

\begin{figure}
\includegraphics[width=0.5\textwidth]{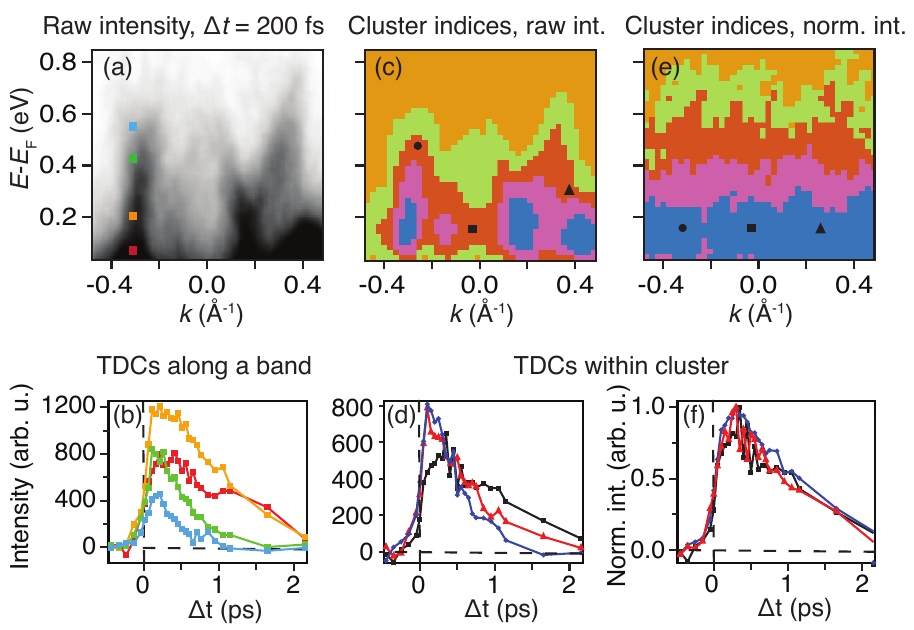}\\
  \caption{(Color online) (a) Excited photoemission intensity at $\Delta t=200$~fs for $h\nu$~=~27.4~eV. (b) Time distribution curves (TDCs) of the photoemission intensity for the regions of interest (ROIs) in panel (a). (c) Result of \km~clustering of the raw TDCs throughout the data set using $k=5$. The $(E, k)$ area is divided into rectangular ROIs and the colour of each ROI corresponds to the cluster index assigned to the corresponding TDC. The markers give the locations of selected TDCs. (d) Selected TDCs from at the locations marked in panel (c). (e) Result of \km~clustering of the same TDCs ($k=5$) after normalising each TDC to a maximum value of 1. (f) Selected normalised TDCs within one cluster, collected at the locations marked in panel (e).}
  \label{lfig:1}
\end{figure}

Fig. \ref{lfig:1}(a) shows the excited photoemission intensity above the Fermi level at $\Delta t=$200~fs, essentially giving an image of the unoccupied electronic structure near the Weyl points (WPs). The dark features correspond to the bands that are unoccupied in equilibrium. As shown in Ref. \cite{short_paper}, these agree qualitatively with density functional theory calculations, especially when $k_{\perp}$ smearing is taken into account. Fig. \ref{lfig:1}(b) shows a few representative TDCs integrated over the rectangular regions of interest (ROIs) marked in panel (a). The TDCs roughly follow a steep band in the unoccupied states and illustrate typical trends in TR-ARPES: a fast initial excitation is followed by a slower decay.  The highest energy TDC shows the fastest decay, consistent with the non-linearity of the Fermi-Dirac distribution. The slower decay of the TDCs closer to the Fermi energy $E_F$ cannot necessarily be described by a (double)-exponential decay. Instead, a more complex behaviour is seen with a plateau, possibly indicating a delayed decay / a continued filling of the state from electrons decaying from higher energies. 

We now divide the entire $(E,k)$ range of the data into ROIs with the same size as in Fig. \ref{lfig:1}(a) and extract the TDC for each ROI. We then apply \km~ to the set of these TDCs using  $k=5$ which is arbitrarily chosen. The result of this clustering is shown in Fig.~\ref{lfig:1}(c) such that each cluster index (1 to 5) is assigned to a color and the initially defined ROIs are coloured according to their cluster index. Note that the cluster indices, and thus the colours, are randomly generated by the \km~ algorithm and do not carry any meaning. Here we choose to order the colours with the highest energy (fastest decay) always having the same colour. This order is arbitrary and carries no meaning but it makes it easier to compare clustering results. The interpretation of  Fig.~\ref{lfig:1}(c) is that regions of the same color contain similar TDCs.  The overall colour landscape in Fig.~\ref{lfig:1}(c) shows a strong similarity to the intensity distribution at $\Delta t=$200~fs in Fig.~\ref{lfig:1}(a). It is easy to understand why: the \km~ algorithm needs to have a metric to calculate the distance of a TDC to a cluster mean value (the cluster centroid) and for this purpose it uses the squared Euclidean distance. This implies a major role of the absolute intensity in a TDC: two TDCs of the same shape but with very different absolute intensities are unlikely to end up in the same cluster, even though their electron dynamics may be similar. Indeed, this is readily seen by an inspection of the three TDCs belonging to the red cluster in Fig.~\ref{lfig:1}(c) and plotted in Fig.~\ref{lfig:1}(d). While the overall intensity of the TDCs is similar, their line shape is rather different. We conclude that clustering the raw TDCs mainly reveals the absolute photoemission intensity at $\Delta t=$200~fs and does not provide much useful new information.

If we are interested in the electron dynamics, and thus the shape of the TDCs, it is more promising to let \km~ operate on suitably \emph{normalised} TDCs. The result of this is shown in Fig.~\ref{lfig:1}(e). Here the TDCs have been normalised to have the same maximum value (1) and a consistent clustering appears over the entire $(E,k)$ range, with cluster distributions resembling horizontal stripes rather than the band positions in  Fig.~\ref{lfig:1}(a). Such horizontal stripes are the expected overall trend with the Fermi-Dirac distribution dictating shorter decay times at higher energies. Fig.~\ref{lfig:1}(f) again shows a selection of TDCs from the same cluster and these look much more similar in shape than in the equivalent plot for the raw data in Fig.~\ref{lfig:1}(d). 

The clustering in Figs.~\ref{lfig:1}(c) and (e) show relatively clear borders between the areas of different cluster indices and a low degree of regions with inter-cluster mixing. This is a non-trivial result because the \km~ algorithm as such is ignorant about the $(E,k)$ locations of the TDCs it operates on. The smooth borders are thus an indication of a suitable ROI definition for the extraction of the TDCs. The ROIs are large enough to generate TDCs of sufficiently low noise for a high-quality clustering result. 

\begin{figure}
\includegraphics[width=0.5\textwidth]{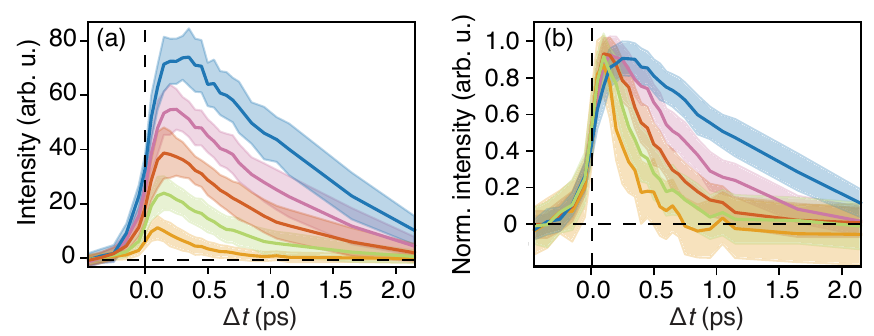}\\
  \caption{(Color online) Cluster centroids for the (a) raw TDCs and (b) normalised TDCs in the \km~clustering of Fig.~\ref{lfig:1}. The colours of the traces are the same as the corresponding cluster colours in Fig.~\ref{lfig:1}. The coloured areas around the TDCs represent the standard deviations.}
  \label{{lfig:2}}
\end{figure}

A useful product of \km~clustering are the so-called cluster centroids. In our case, these are the sum of all TDCs belonging to a cluster divided by the number of TDCs in the cluster. The five cluster centroids for raw and normalised data from  Fig.~\ref{lfig:1} are shown in Figs.~\ref{{lfig:2}}(a) and (b), respectively. Starting with Fig.~\ref{{lfig:2}}(a), the cluster centroid line shape follows the expected trend. The TDCs at the highest energy (lowest cluster index) show the least excitation and the fastest decay. However, this result is not very reliable because of the issue described above: the clustering is not primarily guided by the line shape but rather by the intensity of the TDCs. A more meaningful line shape analysis emerges from the cluster centroids of the normalised TDCs in Fig.~\ref{{lfig:2}}(b). The faster decay at high energies is seen more clearly. In this direct comparison, it is also clear that the rise time during the excitation is identical for all TDCs but there appears to be a delay for the onset of the decay for the centroid closest to the Fermi energy (the blue one).

There are other important aspects when analysing the cluster centroids. Most cluster centroid TDCs have a greatly improved $S/N$ compared to the single ROI TDCs due to the averaging over many TDCs in the cluster. This is also captured by  the standard deviation of the centroid TDCs that are shown in the plot at coloured areas. In the remainder of this paper, these standard deviations are similar in magnitude but they are mostly omitted for clarity of the presentation. The value of the standard deviation is influenced by two factors: the absolute intensity of the signal, giving poorer statistics for TDCs at higher energies, and the number of TDCs in a cluster.  The improved $S/N$ of the cluster centroids is an advantage for a more in-depth line shape analysis but using the cluster centroids for this purpose needs to be done with some care. First of all, one needs to make sure that the TDCs in a cluster are indeed similar to each other as in the case of Fig.~\ref{lfig:1}(f). This is not necessarily the case. After all, the number of clusters $k$ is arbitrarily chosen. A higher $k$ reduces the $S/N$ of the cluster centroids but the TDCs in the clusters will also be more similar to each other, increasing the quality of the parameters extracted from a line shape analysis. Indeed, forming cluster centroids from TDCs with very different line shapes  can lead to erroneous conclusions when inspecting the line shapes of the cluster centroids. A case illustrating this danger are the centroids in Fig.~\ref{{lfig:2}}(a). We have seen that the maximum TDC intensity is more important than the line shape when clustering raw TDCs and so the line shape of a cluster centroid may not be very meaningful. Also in the limiting case of using a small $k$ to describe a data set with large variety, the cluster centroid TDCs loose usable line shape information due to being the average over many differently shaped TDCs. Finally, inspecting the cluster centroids can help to define $k$ for the \km~clustering. Very noisy cluster centroid TDCs and cluster centroid TDCs that are very similar to each other are both indications of $k$ being too high. We have followed this type of guidance to choose $k$ for the cases in this paper. 

Along the same lines, note that in Fig.~\ref{{lfig:2}}(b), the cluster centroids from applying \km~to the normalised TDCs are no longer normalised but have maxima around 0.9. This arises from averaging normalised curves that do not all have the maximum at the same $\Delta t$. A shift in the maximum is not necessarily due to noise. For instance, the highest energy centroid TDC does have a maximum that is clearly shifted to a higher $\Delta t$ with respect to the other TDCs. The deviation of the maximum from 1 may  provide a measure of variation between TDCs classified into one cluster. For a purely visual comparison of subtle line shape differences between the cluster centroid TDC, it can be beneficial to re-normalise these after clustering. 

\begin{figure}
\includegraphics[width=0.5\textwidth]{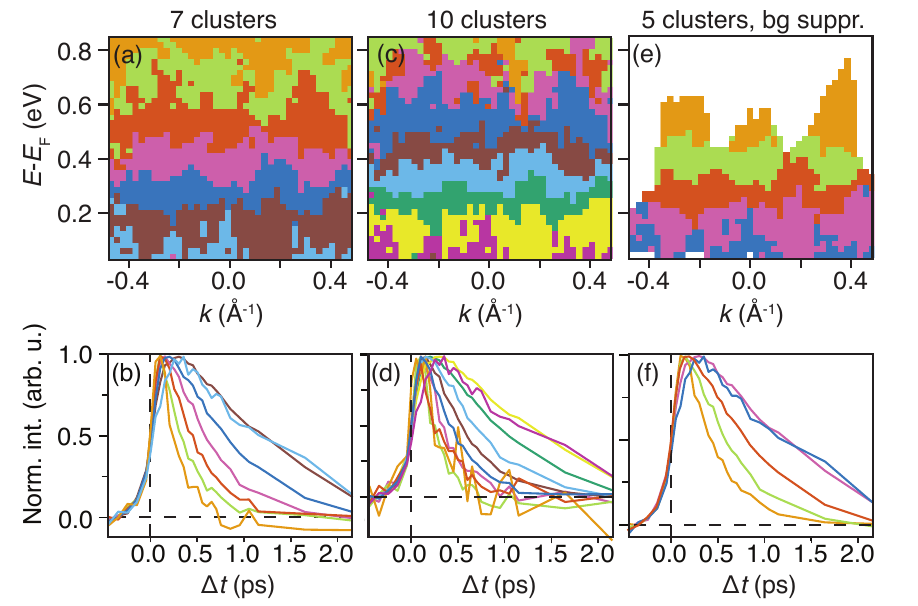}\\
  \caption{(Color online) \km~clustering of data at a single photon energy (27.4~eV). (a) Map of the cluster index distribution for $k=7$. (b) Corresponding cluster centroids. (c), (d) Results for $k$=10. (e) Cluster index distribution for $k=5$ but after excluding the TDCs reaching less than 20\% of the overall maximum intensity in the entire data set for this photon energy. (f) Corresponding cluster centroid TDCs.}
  \label{lfig:3}
\end{figure}

Fig.~\ref{lfig:1}(e) and Fig.~\ref{{lfig:2}}(b) clearly illustrate the overall shorter lifetime of the excited states at higher energies but it is desirable to detect more subtle details in the data. A seemingly obvious way to achieve this is to increase the number of clusters $k$. To illustrate the effect of this, Figs.~\ref{lfig:3}(a) and (c) show the same clustering of normalised TDCs as in Fig.~\ref{lfig:1}(e) but for higher values of $k$ (7 and 10) and Fig.~\ref{lfig:3}(b) and (d) show the corresponding cluster centroids. Increasing the number of clusters does only partly reveal more fine structure, such as a possible $k$-dependence close to $E_\mathrm{F}$. The two lowest energy clusters in Figs.~\ref{lfig:3}(a) and (c) introduce some structure near $E_\mathrm{F}$ that is not seen in Fig.~\ref{lfig:1}(e). There are now two clusters at low energy but the cluster distribution is very similar for $k=7$ and 10, and the centroid TDCs for the lowest energy clusters are also very similar. The more pronounced effect when increasing $k$ is that the additional clusters are mostly distributed in the high energy part of the spectrum. When inspecting the corresponding cluster centroid TDCs, the reason for this becomes clear: at high energies, the excitations are typically weak and there is little signal. The TDCs are thus increasingly dominated by noise. Despite the low signal, the TDCs are still normalised to a maximum of 1, amplifying the randomly distributed noise. The differences between the normalised noisy spectra are then so large that any additional clusters are used to cover this variety. 

In order to reveal details in the electron dynamics, it is thus desirable to exclude noise-dominated TDCs from the clustering. There are two simple ways to achieve this. The first is to restrict the energy region for clustering, cutting off the highest energies that are dominated by noise. This works well but is not shown here. A better approach, retaining more TDCs for clustering, is to inspect the maximum intensity reached in each raw TDC and then to the set a threshold that must be exceeded in order to include a TDC into the set to be clustered. This is illustrated in Fig.~\ref{lfig:3}(e) using again the smaller $k=5$ but excluding ROIs in which the raw TDCs reach less than 20\% of the maximum peak intensity in the entire data set. The result of this approach appears to combine the characteristics of raw intensity clustering in Fig.~\ref{lfig:1}(a) with the constant energy stripes of, e.g., Fig.~\ref{lfig:1}(e). This behaviour can be understood by the clustering still being based on the normalised TDCs, leading to the horizontal stripe pattern, while it can be taken to higher energies into regions where there are bands, and hence there is a resemblance to the outline of the cluster shapes in Fig.~\ref{lfig:1}(c). The approach of restricting the TDCs to be clustered brings out the finer details of the dynamics even with a small number of cluster. This is revealed by comparing Figs.~\ref{lfig:3}(a), (c) and (e). Despite having a smaller total number of clusters (5 vs. 7 and 10, respectively), the clustering in Fig. \ref{lfig:3}(e) reproduces the subtle variations at low energy near $E_\mathrm{F}$. Not surprisingly, the cluster centroids for the two clusters closest to $E_F$ are very similar in Figs.~\ref{lfig:3}(b), (d) and (f).

\section{Data from several photon energies}

\begin{figure}
\includegraphics[width=0.5\textwidth]{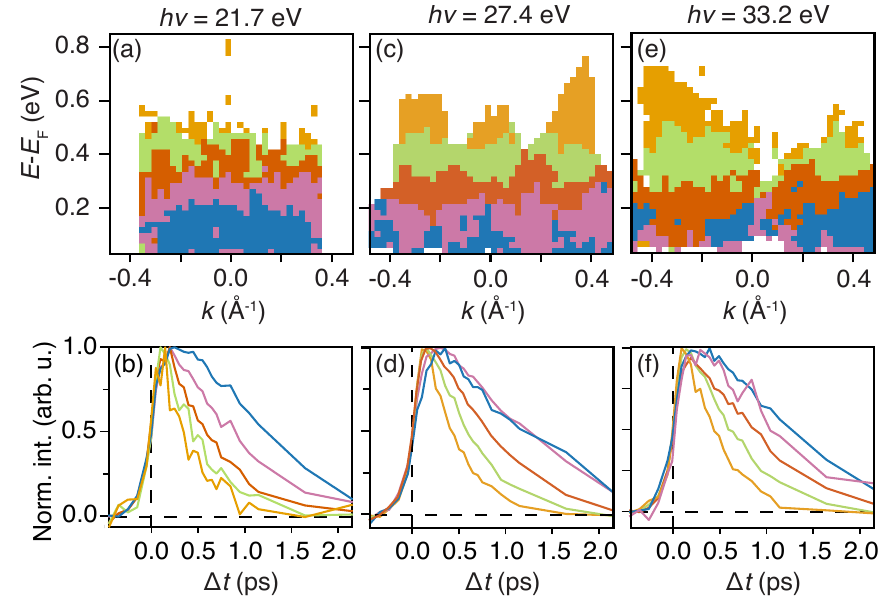}\\
  \caption{(Color online) \km-clustering using a low intensity cutoff and $k=5$ over all three data sets. (a), (c), (e) Cluster index distribution for $h\nu = 21.7$, 27.4 and 33.4~eV, respectively. (b) (d), (f) Corresponding cluster centroid TDCs.}
  \label{lfig:4}
\end{figure}

We now apply the approach introduced in Fig.~\ref{lfig:3}(e) (clustering the entire $(E,k)$ range with an intensity threshold) to all three data sets. The results are shown in Fig.~\ref{lfig:4} as colour maps of the clustering along with TDCs for the cluster centroids. In order to facilitate a detailed comparison between the TDC line shapes, we now re-normalise these cluster centroid TDCs to a maximum value of 1. Fig.~\ref{lfig:4} is  rich in information but the results are not easy to interpret because there is no correspondence of clusters between the different photon energies, i.e., the cluster centroid for a given cluster index / colour is different for each photon energy. One can still draw tentative conclusions by comparing the cluster centroids at similar energies. For example, the green cluster area is approximately at the same energy for $h\nu = 21.7$ and 27.4~eV (Figs.~\ref{lfig:4}(a) and (c)) but the decay time is clearly faster in the green centroid TDC of  Fig. \ref{lfig:4}(b) compared to Figs.~\ref{lfig:4}(d) and (f). Qualitative considerations like this suggest that the dynamics is fastest for $h\nu = 21.7$~eV. This is consistent with Fig.~\ref{lfig:4}(a) showing less excitation to states at high energies, something that could be explained by a very fast decay of such populations, faster than our time resolution. 

For the data taken at $h\nu = 27.4$ and 33.2~eV (Figs.~\ref{lfig:4}(c) and (e)), there is some $k$-dependence near $E-E_\mathrm{F}$ with two clusters falling into this region. As we have seen before, the cluster centroids for these two clusters are very similar for $h\nu = 27.4$~eV. This is not the case for $h\nu = 33.2$~eV.  Finally, the figure illustrates why \km-clustering is a well-suited tool for this type of analysis -- there is a large difference in line shapes in the set of cluster centroid TDCs and it would be very challenging to fit the data using a single line shape model. 

A consistent classification of the electron dynamics throughout the entire data set can be achieved by taking \emph{all} the TDCs in the entire data set as input for \km~clustering. In order to allow for photon energy-dependent photoemission matrix element variations, the data set at each photon energy is normalised to the same maximum value before extracting the TDCs and then a common intensity threshold is defined to exclude low-intensity TDCs from clustering (lower than 20\% of the absolute intensity maximum). The results of this approach are shown in Fig.~4(a) and (b) of Ref. \cite{short_paper} as cluster maps and cluster centroid TDCs. Now the colours across the cluster maps can be compared on equal footing since they stand for the same cluster index throughout the data set and the faster electron dynamics for  $h\nu = 21.7$~eV becomes evident by visual inspection, as discussed in Ref.~\cite{short_paper}.

\begin{figure}
\includegraphics[width=0.3\textwidth]{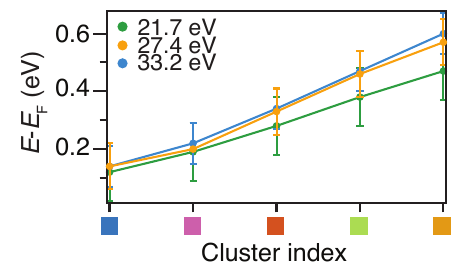}\\
  \caption{(Color online) Mean energy of the cluster positions for a given cluster index (colour) and for the three different photon energies. The lines between the points are a guide to the eye. }
  \label{lfig:5}
\end{figure}

Here we present a more quantitative analysis of the decay time. Fig.~\ref{lfig:5} shows the mean energies for the clusters at the three photon energies with the clusters ordered by energy. These mean energies are calculated from the energies of the ROIs that have been assigned to a particular cluster. The energy for which a certain behaviour / TDC shape is found is almost identical for $h\nu = 27.4$ and 33.4~eV but it is different for $h\nu = 21.7$~eV. For a given cluster label, the mean energy for $h\nu = 21.7$~eV is consistently lower than for the two other photon energies, implying that that the same dynamics takes place at a lower energy or, in other words, that the overall dynamics is faster. The difference between $h\nu = 21.7$~eV and the other two photon energies is especially large at high energies. 

As pointed out in Ref.~\cite{short_paper}, the faster electron dynamics at $h\nu = 21.7$~eV can be explained by an inspection of the bulk Fermi surface (Fig.~1(a) in Ref.~\cite{short_paper}). When using $h\nu = 21.7$~eV, one probes states towards the $A-L-H$ plane of the BZ and this is where the bulk Fermi surface is located, in contrast to the $\Gamma-M-K$ plane that is (approximately) explored  with $h\nu=$27.4 and 33.4~eV and does not have any Fermi surface segments nearby (apart from the yellow ``cigars'' in Fig. 1(a) of Ref. \cite{short_paper}, but these might be an artefact of the calculation, as discussed there). The metallic states thus render the dynamics at $h\nu = 21.7$~eV faster, as one would naively expect. Indeed, one might even ask why the difference between the ``metallic'' and ``insulating'' regions of the BZ is not even more pronounced.  It is clear that a very fast decay in the metallic region might not be observable due to the $k_\perp$ smearing effect discussed in detail in Ref.~\cite{short_paper}. After all, the quantitative comparison in Fig.~2(c) of Ref.~\cite{short_paper}  suggests that this smearing stretches over about 15\% of the BZ size, so that the slow decay in the insulating part of the BZ would mask out the fast decay near the $A-L-H$ plane. 

The data in Fig.~\ref{lfig:4} suggests that there may be some subtle differences in the $k$-dependence of the TDCs close to $E_F$ between the data sets collected at $h\nu=27.4$ and 33.4~eV, i.e. between the two scans approximately representing the cuts in Fig.~1(b) of Ref.~\cite{short_paper}. Indications of such differences were indeed seen in Fig.~4 of Ref.~\cite{short_paper} by performing clustering on the concatenated TDCs from both data sets. In the following, we discuss the reasoning behind this approach and we confirm the findings of \km~analysis by a more conventional inspection of TDCs in different regions of interest. 

When clustering the concatenated TDCs from two (or several) photon energies, what can be learned from the result? In the simplest case, the dynamics would be the same for all ROIs with corresponding $k$ and $E$ at both photon energies and a concatenated TDC would just show the same dynamics twice. The clustering map would then be identical for each individual photon energy and for the concatenated TDCs. ROIs that show a different dynamics for the two photon energies could still end up in one cluster, as long as the difference is always the same. On the other hand, changes in the difference could result in different cluster assignments.

\begin{figure}
\includegraphics[width=0.5\textwidth]{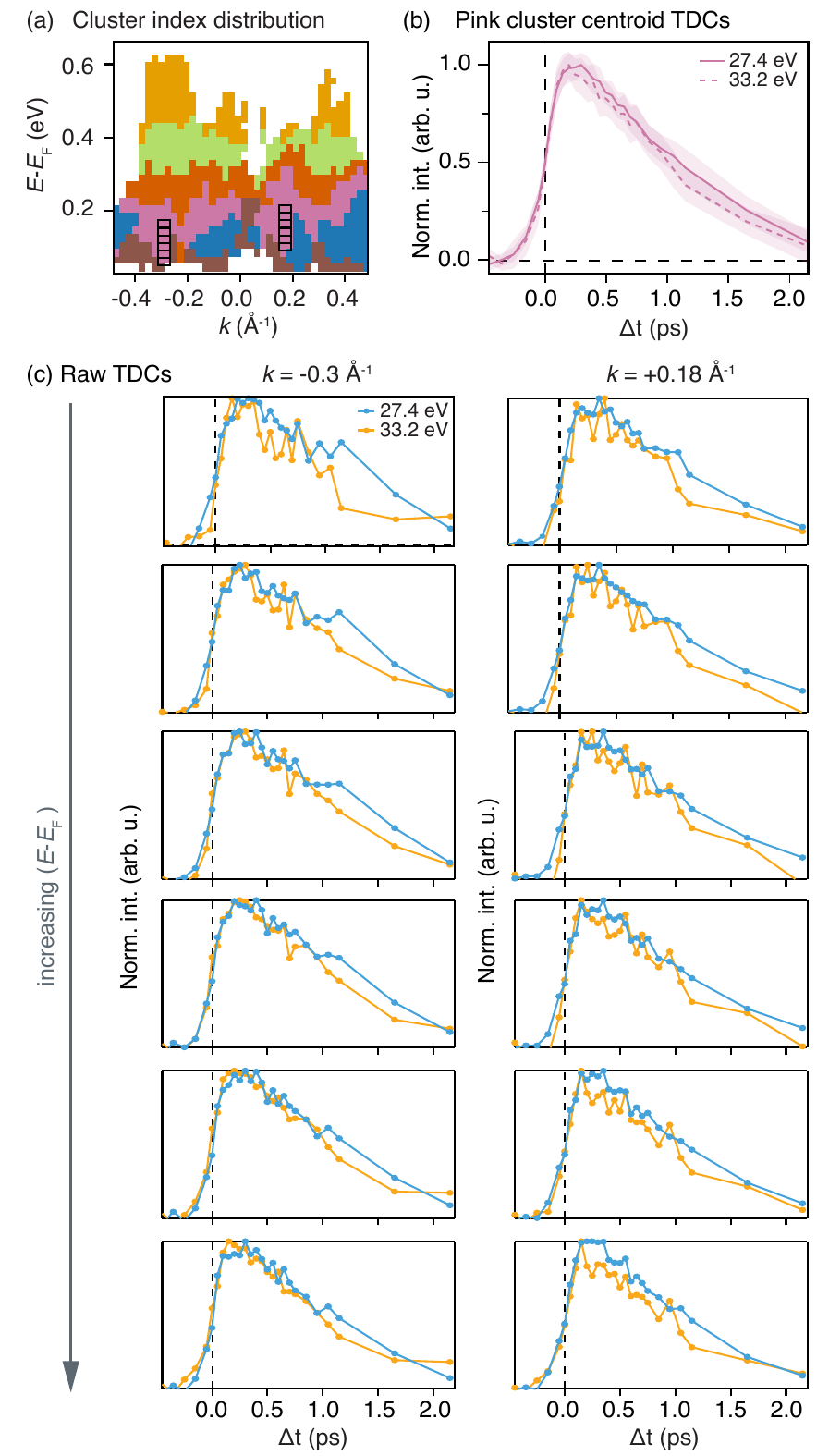}\\
  \caption{(Color online) (a) Clustering result for the concatenated TDC for $h\nu=27.4$ and 33.4~eV (identical to Fig. 4(c) in Ref. \cite{short_paper}). ROIs for further analysis in panel (c) are marked by black rectangles. (b) Centroids for the purple cluster (iii) in panel (a), with the TDCs for the two photon energies plotted on top of each other.  (c) TDCs for the two photon energies for the ROIs in panel (a).  }
  \label{lfig:6}
\end{figure}

The clustering result of the concatenated TDCs for $h\nu=27.4$ and 33.4~eV is shown in Fig.~\ref{lfig:6}(a). All the cluster centroids are given in Fig.~4(e) of Ref.~\cite{short_paper}, such that the concatenated TDCs are split up again and the TDCs for the two photon energies are compared to each other. For most of the clusters, the two TDCs are essentially identical but there are three exceptions: cluster indices (i),  (iii) and (vi), with the difference for cluster (iii) being most pronounced. The two centroid TDCs for cluster (iii) are shown again here in Fig.~\ref{lfig:6}(b). For the longer $\Delta t$ values, the TDC for $h\nu = 27.4$~eV tends to have less intensity than the TDC for $h\nu = 33.2$~eV. This could indicate that either the decay is the same and the maximum is reached earlier or that the decay is slightly faster. The differences are not statistically significant (the two curves are within one standard deviation from each other) but they are evident when compared to the other curves in Fig.~4(e) of Ref.~\cite{short_paper}. 

Fig.~\ref{lfig:6}(c) shows two sets of normalised TDCs from cluster (iii), in regions where it is possible to track TDCs in vertically stacked ROIs (ROIs for the same $k$), as shown in Fig.~\ref{lfig:6}(a). In order to improve the statistics here, the ROIs are twice as large as those used for clustering. The trend of the $h\nu = 27.4$~eV TDCs to show a higher intensity  than for the $h\nu = 33.2$~eV TDCs at long $\Delta t$ is clearly visible throughout the data set. Indeed, the differences appear even clearer than in Fig.~\ref{lfig:6}(b), presumably because of the averaging effect in the cluster centroids. 

The difference between the two photon energies is very subtle and not straight-forward to interpret. It is curious that it is mostly found in parts of the data set and in an energy range fairly high about the WPs. The tendency for a faster decay at $h\nu = 33.2$~eV could be tentatively ascribed to the slightly smaller distance (in $k_\perp$) to the metallic part of the BZ. This can clearly be seen in Fig.~3(b) of Ref.~\cite{short_paper} where $k_\perp$-smearing leads to an intensity of spectral intensity in the projected gap around $E_\mathrm{F}$ which is absent for  $h\nu = 27.4$~eV.

\section{Conclusion}

We have applied \km~clustering of TDCs to ARPES data taken as a function of energy, $k$ along one direction, pump-probe time delay and probe photon energy, so as to explore the electron dynamics in the entire 3D BZ of the Weyl semimetal PtBi$_2$. When applied to TDC line shapes rather than to absolute intensities, this approach can reveal subtle trends in the complex data. In particular, \km~clustering allowed us to find a faster dynamics in the parts of the BZ hosting the material's Fermi surface, as well as subtle TDC line shape differences between the BZ region hosting the Weyl points and a nearby region. The most pronounced changes in TDCs typically appear as a function of energy, simply due to the strong non-linearity of the Fermi-Dirac distribution, and this is reflected in the ease with which \km~clustering can distinguish between the different line shapes of energy dependent TDCs, also between photon energies. 

It is not \emph{a priori} clear that \km~clustering is a suitable tool for the problem at hand. After all, \km~is designed to cluster objects into distinct classes whereas the type of data we are interested in represents a more continuous variation. For instance, the typical decay time for excited electrons decreases for higher energies in a continuous way. On the other hand, \km~is routinely used for similarly continuous problems, for example in colour quantisation when compressing images \cite{Celebi:2011aa}. 

The most important advantage of applying \km~here is that it enables us to find trends in a multi-dimensional data set, excluding human bias in, e.g., selecting specific ROIs to perform a more detailed analysis on. It is clear that this advantage will increase in importance for data sets with an even higher dimensionality, for instance when varying other experimental parameters such as the pump photon energy, fluence or light polarisation.

\begin{acknowledgments}
This work was supported by the Independent Research Fund Denmark  (Grant No. 1026-00089B). Access to Artemis at the Central Laser Facility was provided by STFC (Experiment Number 23120004). SA acknowledges DFG through AS 523/4-1.
\end{acknowledgments}

%

\end{document}